\documentclass[review]{revtex4}
\usepackage{lineno,hyperref}
\usepackage{amsmath}
\usepackage{amssymb}
\usepackage{graphicx}
\usepackage{float}
\modulolinenumbers[10]
\usepackage{mathtools}
\usepackage{xcolor}
\usepackage{csquotes}

\usepackage{xpatch}
\makeatletter
\xpatchcmd{\maketitle}{\if@twoside\next@tpage}{\iffalse}{}{}
\makeatother

\begin{document}

\preprint{APS/123-QED}
	
\title{Late time dynamics of  $f(R, T, R_{\mu\nu}T^{\mu\nu})$  gravity}

\author{Maryam Aghaei Abchouyeh$^{1}$}
\email{m.aghaei@ph.iut.ac.ir}
\author{Behrouz Mirza$^{1}$   }
\email{b.mirza@iut.ac.ir}
\author{Parisa Shahidi$^{1}$}
\author{Fatemeh Oboudiat$^{1}$}
\email{f.oboudiat@ph.iut.ac.ir}

\affiliation{$^{1}$ Department of Physics, Isfahan University of Technology, Isfahan 84156-83111, Iran}

\begin{abstract}
Dynamical behavior and future singularities of $f(R, T,R_{\mu\nu}T^{\mu\nu})$ gravitational theory are investigated. This gravitational model is a more complete form of the $f(R,T)$ gravity which can offer new dynamics for the universe. We investigate this gravitational theory for the case $f = R + \alpha R_{\mu\nu}T^{\mu\nu}$ using the method of autonomous dynamical systems and by assuming an interaction between matter and dark energy. The fixed points are identified and the results are consistent with standard cosmology and show that for small $\alpha$, the radiation dominated era is an unstable fixed point of the theory and the universe will continue its procedure toward matter era which is a saddle point of the theory and allows the evolution to dark energy dominated universe. Finally the dark energy dominated epoch is a stable fixed point and will be the late time attractor for the universe. We also consider future singularities for the two $f = R + \alpha R_{\mu\nu}T^{\mu\nu}$ and $f = R +\alpha RR_{\mu\nu}T^{\mu\nu}$ cases and for $w = 0,\dfrac{1}{3},1$ and $-1$. Our results show that for the case of $f = R + \alpha R_{\mu\nu}T^{\mu\nu}$, the future singularities of the universe will happen in the same condition as do for the Einstein-Hilbert FRW universe. However, a new type of singularity is obtained for $f = R +\alpha RR_{\mu\nu}T^{\mu\nu}$ that is captured by $t\rightarrow t_s;\  a \rightarrow a_s;\  \rho\rightarrow \infty;$ and $\  |p| \rightarrow 0$.
\end{abstract}

\maketitle

\section{Introduction}
\label{intro}
Recent observational data from Type I Supernova and CMB are the most important evidences confirming accelerated expansion of our universe\cite{1,1a,1b,1c}. Two mechanisms may be invoked to explain this expansion. The first is based on the assumption of the existence of ``dark energy" within the framework of General Relativity (GR) with a negative pressure in which the equation of state $p /\rho<0$ holds, where $\rho$ represents the energy density and $p$ is the pressure of the universe. On the other hand , observation has revealed that $w=p/\rho$ is slightly less than $- 1$, which refers to the phantom dark energy universe \cite{2,2a,2b,2b2,2c}. There is another type of dark energy called  quintessence and expressed by the equation of state $w = p/\rho >-1$.

\textcolor{black}{Many different forms of dark energy has been proposed in the literature, but it is still an exotic form of energy with many ambiguities surrounding it. Thus, it is getting more popular in the cosmology studies to explore modified gravitational theories as alternatives to dark energy.}
The most popular modifications to the GR theory are \textcolor{black}{$f(R)$, $f(R,T)$ and $f(T)$ theories \cite{S8,3a,S7,S14,S16,S5,S18,S17,3b,fixfrt,sc1,sc2,sc3} in which $R$ is the Ricci scalar and $T$ is the trace of energy-momentum tensor}. Other types of generalizations have also been investigated over the past decades with different purposes from applications in cosmology as alternatives to the dark energy, to modifying the early universe dynamics to remove the big bang singularity.
	
	One of the known extensions to GR theory is the familly of gravitational theories called $f(R,T)$ theory which includes non-minimal couplings between matter and curvature. The recent book by Harko and Lobo contains a very good introduction for these theories and their motivation and applications \cite{S10}.  

	Different properties of $f(R,T)$ gravitational theory has been studied and explained during the past decade \cite{S3,S4,S5,S9,S6,S10}. 
	It is known that there exist $f(R,T)$ gravitational models that can explain the cosmological epochs and the late time accelerated universe \cite{fixfrt}. In addition it is possible to explain the observational scale depended growth of scalar perturbations for sub-hubble limit in this theory \cite{S2}. However, in case of traceless energy-momentum tensor ($T=0$), the field equations of $f(R,T)$ theory will reduce to that of $f(R)$ theory. Therefore all the non-minimal couplings between matter and geometry will disappear. \textcolor{black}{This reason makes a motivation to suggest a more general coupling between matter and geometry as $R_{\mu \nu}T^{\mu\nu}$ where $R_{\mu \nu}$ and $T^{\mu \nu}$ are Ricci tensor and energy- momentum tensor, respectively. Including such a coupling in a gravitational theory will keep the non-minimal coupling between matter and geometry even in the case of $T=0$. For instance, this feature is important when the electromagnetic tensor is included in the matter Lagrangian. $f(R, T, R_{\mu \nu} T^{\mu \nu})$ theory is one of the recent proposals for modified gravitational theories and was first proposed as an extension for Ho\v{r}ava-Lifshitz theory with dynamical Lorentz symmetry breaking\cite{S10,5a,S1,S13,S12,S11}. }

	Odintsov and S\'{a}ez-G\'{o}mez have studied the stability of matter part of the theory, the natural occurrence of $\Lambda CDM$, the reproduction of de-sitter universe where the non-constant fluid is present, and the conditions to satisfy the continuity equation.
	
	Almost simultaneously another group studied this theory using another approach \cite{5a}. The authors purposed $f(R, T, R_{\mu \nu}T^{\mu \nu})$ gravity as an extension for $f(R,T)$ gravitational theory.\\
	
	They studied the extra force due to the coupling between matter and Ricci tensor, on the moving massive particle in this gravitational theory. They also investigated the cosmological implications of the theory for both conserved energy-momentum tensor and the non-conserved one, the stability and the Newtonian limit of the theory. \textcolor{black}{Other aspects of $f(R,T,R_{\mu \nu}T^{\mu \nu})$ gravitational theory have been studied in literature \cite{S13,S12,S11}.}\\
	\indent

	\textcolor{black}{Generally different non-minimal matter-curvature coupling in a theory allows for energy exchange between these two sectors. If the coupling is between the scalars such as $R$ and $T$, we can inspire new dynamics associated with scalar part. In the case of  $f(R, T, R_{\mu\nu}T^{\mu\nu})$ theory the coupling between matter and curvature is in tensorial part so  potentially there would be new dynamics, new mechanism for late time acceleration and new types of  future singularities for the universe. It is what motivates us here to study the dynamical behaviors and future singularities  of this theory.}\\
	It is expected for dynamical behavior of a proposed gravitational theory to keep the time ordering of standard cosmology epochs in which there is a radiation dominated epoch followed by a matter domination era and finally a dark energy universe. To study the dynamics of this theory, the field equations of $f(R, T, R_{\mu\nu}T^{\mu\nu})$ theory are considered using the autonomous dynamical system method \cite{frtheory,fixfr,fixft1,fixft2,fixft3,fixft4,fix,fixfrt}. We first find the fixed points of the theory for the universe that contains radiation, matter and dark energy. Studying the stability of these fixed points, we show that this gravitational theory can keep the time ordering and the stability of the different cosmological epochs.\\
		We also investigate two special case of the theory for the four different conditions of $w=-1$, $w=1/3$,  $w=1$, and $w=0$ representing dark energy,  radiation, high cosmological energy density, and  dust matter universe, respectively, to check the future singularities of the universe. It is interesting to investigate different classifications of the future singularities in modified gravities \textcolor{black}{because it makes an opportunity for us to understand the properties and weak points of different gravitational theories. This will help us to classify the behaviors and singularities of extended gravitational theories and to introduce viable consistent models.}
	In general, the singularities of the universe are defined by divergence or vanishing of certain cosmological parameters such as: Hubble parameter $H$, scale factor $a(t)$, energy density $\rho$, or pressure $p$.\\	
	\\
\indent	Our results show that, although for small matter-curvature coupling and small interaction between matter and dark energy, the dynamical evolution of the present model is compatible with the standard cosmological expectation, the universal classification of the singularities can not be hold in this theory. The reason is a new type of singularity that is predictable for the future of the universe which is described by this gravitational theory.

	\textcolor{black}{This paper is organized as follows. The $f(R)$, $f(R,T)$  and $f(R, T, R_{\mu\nu}T^{\mu\nu})$ theories are reviewed briefly in Section \ref{II}. In Section \ref{III}, we investigate the $f(R, T, R_{\mu\nu}T^{\mu\nu})$ theory using  the autonomous dynamical system method. In Section \ref{IV}, the future singularities of $f=R+\alpha R_{\mu \nu}T^{\mu \nu}$ are studied for the different cases of  $w=-1,0, \dfrac{1}{3},  1$. Section \ref{V} investigates the future singularities for the case $f=R+\alpha RR_{\mu \nu}T^{\mu \nu}$ to discover a new type of  singularity. Finally section \ref{V1} presents some numerical analysis of this theory using Cosmographic parameters to find the values of free parameters of this theory.}

	\section{$f(R, T, R_{\mu\nu}T^{\mu\nu})$ Gravity}
	\label{II}
	This section describes the general features of the $f(R)$, $f(R,T)$ and $f(R, T, R_{\mu\nu}T^{\mu\nu})$ gravitational theories. In these theories, the action is a function of the \textit{Ricci, Ricci scalar and energy momentum tensor}, thus in the presence of matter is written as follows \cite{frtheory}:
	\begin{equation}
	\label{a}
	I=\int d^4x\sqrt{-g}\hspace{0.1cm}\Big[\dfrac{f(R)}{2\kappa}+\mathcal{L}_m\Big],
	\end{equation}
	
	\noindent where, \textcolor{black}{$\kappa=8\pi G$, $\mathcal{L}_m$ is the matter Lagrangian density and $g$ is the determinant of the metric}.
	Variation of Eq. (\ref{a}) with respect to the metric yields the equation of motion as follows:

	\begin{equation}
	\label{b}
	\kappa T_{\mu\nu}^{matter}=f'(R)R_{\mu\nu}-\frac{1}{2}g_{\mu\nu}f(R)+g_{\mu\nu}\Box f'(R)-\nabla _{\mu} \nabla _{\nu}f'(R).
	\end{equation}

	We assume a flat isotropic space time with the following FRW metric:
	
	\begin{equation}
	\label{c}
	ds^2=-dt^2+a^2(t)\gamma_{ij}dx^i dx^j,
	\end{equation}
	
	\noindent where, $a(t)$ is the scale factor and $\gamma_{ij}dx^i dx^j=dr^2+r^2d\Omega^2$ is the spatial part of the metric. The field equations of the action in Eq. (\ref{a}) then turn to the following form:
	
	\begin{eqnarray}
	\label{d}
	&&H^2(t)=\frac{\kappa}{3f'(R)}(\rho+\rho_{c}),\\
	&&\dot{H}=\frac{-\kappa}{2f'(R)}(\rho+p+\rho_{c}+p_c),
	\end{eqnarray}
	
	\noindent in which, the Hubble parameter is $H(t)=\frac{\dot{a(t)}}{a(t)}$; $p$ and $\rho$ are matter pressure and energy density, respectively; and $p_c$ and $\rho_{c}$ are curvature pressure and curvature energy density, respectively.

	$f(R,T)$ gravity is an extension to $f(R)$ theory where $T$ is the trace of energy-momentum tensor, and the non-minimal coupling between matter and geometry is included in the theory \cite{S7}:
	
	\begin{equation}
	\label{frt}
	S=\frac{1}{2\kappa}\int
	f\left(R,T\right)\sqrt{-g}\;d^{4}x+\int{L_\mathrm{m}\sqrt{-g}\;d^{4}x}.
	\end{equation}
	
	Varying the action in Eq.\eqref{frt} with respect to the metric, we will have the field equation of $f(R,T)$ as bellow:
	
	\begin{eqnarray}\label{field}
	f_{R}\left( R,T\right) R_{\mu \nu } - \frac{1}{2}
	f\left( R,T\right)  g_{\mu \nu }
	+\left( g_{\mu \nu }\square -\nabla_{\mu }\nabla _{\nu }\right)
	f_{R}\left( R,T\right) =\nonumber\\
	8\pi T_{\mu \nu}-f_{T}\left( R,T\right)
	T_{\mu \nu }-f_T\left( R,T\right)g^{\alpha \beta }\frac{\delta T_{\alpha \beta
	}}{\delta g^{\mu \nu}}.
	\end{eqnarray}
	
	It is obvious that for $T=0$, we will regain the $f(R)$ gravity field equation and all the non-minimal matter-eometry coupling will disappear.\\ 
	This is why we need to introduce a more general form of $f(R,T)$ theory with a non-minimal coupling between matter and geometry in tensorial sector, and can be addressed as $f=f(R, T, R_{\mu\nu}T^{\mu\nu})$, where $R$ is the Ricci Scalar, $T^{\mu\nu}$ represents the energy-momentum tensor, and $T=T_{\mu}^{\mu}$, the action can be written as follows \cite{S1,5a}:
	
	\begin{equation}
	\label{action}
	S=\frac{1}{2\kappa}\int d^4x\sqrt{-g}\ f(R,T,R_{\mu\nu}T^{\mu\nu})+\int d^4x\sqrt{-g}\ \mathcal{L}_m.
	\end{equation} 
	
	\textcolor{black}{Using action in Eq.\eqref{action}, one can obtain the general field equations consistently as bellow:}
	
	\begin{eqnarray}
	\label{eq203}
	&&(f_R-f_{RT}L_m)G_{\mu\nu}+[\Box
	f_R+\dfrac{1}{2}Rf_R-\dfrac{1}{2}f+f_TL_m+\dfrac{1}{2}\nabla_\alpha\nabla_\beta(f_{RT}T^{
		\alpha\beta})]g_{\mu\nu}-\nonumber\\	
	&&\nabla_\mu\nabla_\nu f_R+\dfrac{1}{2}\Box(f_{RT}T_{\mu\nu})
	+2f_{RT}R_{\alpha(\mu}T_{\nu)}^{~\alpha}-\nabla_\alpha\nabla_{(\mu}[T^{\alpha}_{
		~\nu)}f_{RT}]\nonumber\\
	&&-(f_T+\dfrac{1}{2}f_{RT}R+8\pi
	G)T_{\mu\nu}-2(f_Tg^{\alpha\beta}+f_{RT}R^{\alpha\beta})\dfrac{\partial^2
		L_m}{\partial g^{\mu\nu}\partial g^{\alpha\beta}}=0.
	\end{eqnarray}
	
	\textcolor{black}{The dynamical behavior and the pressure and energy density of the universe may now be checked} using the filed equations thus obtained from Eq.\eqref{eq203}.\\
	\textcolor{black}{The most general form of the Lagrangian which can be described by Eq. \eqref{eq203} is of the form $R+a T+bRT+cR^2+d R_{\mu \nu} T^{\mu \nu}$. But as we are interested in studying the effect of the coupling between matter and geometry on the singularity and dynamics of the theory, we will isolate the terms including this coupling to have control on their impacts on the dynamics of the theory.}

	\noindent We investigate the cases knowing that the energy-momentum tensor is not conserved in this gravitational theory.\\
	It should be noted that in vacuum our theory is equivalent to general relativity. Therefore this theory is consistent with the results obtained by LIGO (GW170817 and GRB170817) \cite{abbott}.

	\section{Dynamical analysis of $f=R+\alpha R_{\mu \nu}T^{\mu \nu}$ model}
	\label{III}
	The dynamical system analysis have been applied to different modified gravitational theories, such as $f(R)$ and $f(R,\mathcal{G})$ theories \cite{sc4,sc5,sc6} which contain curvature invariants. Here in our model, instead of having further curvature invariants, matter and curvature are coupled to each other as an invariant. In this section, we study the autonomous dynamical systems for the $f=R+\alpha R_{\mu \nu}T^{\mu \nu}$ model where, $\alpha$ is a constant. Assuming $f=R+\alpha R_{\mu \nu}T^{\mu \nu}$ in the action, and using Eq.\eqref{eq203}, the field equations are given as follows \cite{5a}:
	
	\begin{eqnarray}
	\label{first}
	&&G_{\mu\nu}+\alpha\Bigg[2R_{\sigma(\mu}T^\sigma_{~\nu)}-
	\dfrac{1}{2}R_{
		\rho\sigma}T^{\rho\sigma}g_{\mu\nu}-\dfrac{1}{2}RT_{\mu\nu}
	-\dfrac{1}{2}\left(2 \nabla_\sigma\nabla_{(\nu}
	T^\sigma_{~\mu)}-\Box
	T_{\mu\nu}-\nabla_\alpha\nabla_\beta T^{\alpha\beta}g_{\mu\nu}\right)\nonumber\\
	&&-G_{\mu\nu}L_m-2R^{\alpha\beta}\dfrac{\partial^2 L_m}{\partial g^{\mu\nu}\partial
		g^{\alpha\beta}}\Bigg]-
	8\pi GT_{\mu\nu}=0,
	\end{eqnarray}
	
	where for isotropic and homogeneous FRW universe, the cosmological field equations will be obtained as bellow:	
	
	\begin{eqnarray}
	\label{friedmann}
	&&3H^2=\frac{\kappa}{1-\alpha \rho}\rho+\frac{3}{2}\frac{\alpha}{1-\alpha \rho}H(\dot{\rho}-\dot{p}),\label{e}\\
	&&2\dot{H}+3H^2=\frac{2\alpha}{1+\alpha p}H\dot{\rho}-\frac{\kappa p}{1+\alpha p}+\frac{1}{2}\frac{\alpha}{1+\alpha p}(\ddot{\rho}-\ddot{p}).\label{f}
	\end{eqnarray}
	
	\indent
	where, $ \rho =\rho _{m}+\rho _r+\rho _{d}$ and $ p=p_m+p_r+p_d$ and for $\alpha=0$, we regain the standard Friedmann equations. The indices $r$, $m$, $d$ stand for radiation, matter, and dark energy, respectively. Conservation of the energy-momentum tensor does not hold for $f(R, T, R_{\mu\nu}T^{\mu\nu})$ gravity and the explicit continuity equation reads as follows:
	
	\begin{eqnarray}
	&&\frac{d}{dt}(\rho_m+\rho_r+\rho_d)\nonumber\\
	&&+3H(\rho_m+\rho_r+\rho_d+p_m+p_r+p_d)=0.
	\label{contin1}
	\end{eqnarray}
	
	\subsection{The set of dynamical equations}
	
	For a dynamical system analysis of this theory, Eq. \eqref{contin1} needs to be simplified. Hence, the following relations are assumed:
	
	\begin{eqnarray}
	&&\dot{\rho}_r+3H(\rho_r+p_r)=0,\nonumber\\
	&&\dot{\rho}_m+3H(\rho_m+p_m)=-Q, \label{contin2}\\
	&&\dot{\rho}_d+3H(\rho_d+p_d)=Q. \nonumber
	\end{eqnarray}
	
	Eq. (\ref{contin2}) assumes that $Q$ represents the interaction between matter and dark energy and that energy $Q$ is exchanged between typical matter and dark energy. This causes the dark energy portion of the universe to increase as the matter decreases in the universe. It is clear from Eq. (\ref{contin2}) that $Q$ has a dimensions of $\dot{\rho}$ or $H\rho$. Thus, terms like $H\rho_m$, $H\rho_d$, or a combination thereof maybe substituted for $Q$. For our present purpose, we choose $Q=\gamma H\rho_m$, where $\gamma$ is a constant. Assuming that $\omega_r=\frac{1}{3}$, $\omega_m=0$, and $\omega_d=-1$, the continuity equations will take the following forms:
	
	\begin{eqnarray}
	\label{conserv}
	&&\dot{\rho}_r+4H\rho_r =0,\nonumber\\
	&&\dot{\rho}_m+3H\rho_m=-\gamma H\rho_m,\label{contin33}\\
	&&\dot{\rho}_d=\gamma H\rho _m. \nonumber
	\end{eqnarray}
	
	Using Eq. \eqref{contin33}, the dimensionless parameters may be defined as follows:
	
	\begin{eqnarray}
	\Omega _m&=&\frac{\kappa \rho _m}{3H^2},\nonumber\\
	\Omega_r&=&\frac{\kappa \rho _r}{3H^2},\nonumber\\
	\Omega_d&=&\frac{\kappa \rho _d}{3H^2},\label{params}\\
	x&=&\alpha \rho _m,\nonumber\\
	y&=&\alpha \rho _r, \nonumber\\
	z&=&\alpha \rho _d. \nonumber
	\end{eqnarray}
	
	The Eqs. \eqref{e}-\eqref{f} in terms of the dimensionless variables will become:
	
	\begin{eqnarray}
	1&=&\Omega_r + \Omega _m + \Omega _d + \dfrac{1}{2}(x (\gamma - 1) + \dfrac{- 2}{3}y) + z, \nonumber\\
	x&=& \dfrac{y \Omega _m}{ \Omega _r }, \nonumber\\
	z&=& \dfrac{y \Omega _d}{ \Omega _r }, \label{motion2}
	\end{eqnarray}
	
	\noindent and:
	\begin{equation}
	\label{dot}
	\dfrac{\dot{H}}{H^2}=\dfrac{A_1}{B_1},
	\end{equation}
	where
	
	\begin{eqnarray}
	&&A_1=\Omega_r(-16+24\Omega_r+3\Omega_m^2(3+\gamma^2)+\nonumber\\
	&&\Omega_m (34\Omega_r-3(3+\Omega_r)\gamma+3(-1+\Omega_r)\gamma^2)+\nonumber\\
	&&\Omega_d(-8\Omega_r+3\Omega_m(-6+\gamma(3+\gamma))),\nonumber\\
	\nonumber\\
	&&B_1=12\nonumber\Omega_d^2+\Omega_r(8-12\Omega_r)+3\Omega_m^2(-3+\gamma)+\nonumber\\
	&&\Omega_m(3-21 \Omega_r++3(1+\Omega_r)\gamma)+\Omega_d (3\Omega_m(1 + \gamma)).\nonumber
	\end{eqnarray}
	
	It is obvious that the variables $x$ and $z$ depend on $y$. It is now possible to construct the autonomous dynamical system using  the continuity equations in \eqref{contin33} and the Eqs in \eqref{params}, \eqref{motion2}. Thus we will make the derivatives of the variables with respect to the number of e-folding as follows:
	
	\begin{eqnarray}
	\frac{d\Omega_m}{dN}&=& (-3\Omega_m-\gamma\Omega_m)-2\Omega_m\dfrac{\dot{H}}{H^2}, \nonumber\\
	\frac{d\Omega_r}{dN}&=&(-4\Omega_r)-2\Omega_r\dfrac{\dot{H}}{H^2},\nonumber\\
	\frac{d\Omega_d}{dN}&=&\gamma\Omega_m-2\Omega_d\dfrac{\dot{H}}{H^2},\nonumber\\
	\frac{dx}{dN}&=&(-\gamma - 3)x\label{system},\nonumber\\
	\frac{dy}{dN}&=&-4y,\nonumber\\
	\frac{dz}{dN}&=&\gamma x,
	\end{eqnarray}
	
	where, $N=\int Hdt$. These equations describe the dynamics of the energy portions of the universe. Considering  Eqs. (\ref{e}), (\ref{f}) and (\ref{contin33}) and  the dependency of $x$ and $z$ to $y$ (Eq. \ref{motion2}), we conclude that only two variables out of the seven variables, $\Omega_m,\hspace{0.1cm} \Omega_r,\hspace{0.1cm} \Omega_d,\hspace{0.1cm}x,\hspace{0.1cm} y,\hspace{0.1cm} z,\hspace{0.1cm} \dfrac{\dot{H}}{H}$,  are independent. So it is needed to find the two independent ones. To obtain the two independent variables, we should notice that as $x$ and $z$ are have dependency to $y$, we may just keep $y$ as independent one. From the Eqs. (\ref{motion2}) and Eq. (\ref{dot}), $y$ and $\dfrac{\dot{H}}{H^2}$ can be written in terms of three other ones. Thus up to  here we have three variables that we should omit one of them.  To achieve this purpose the derivative of the first equation of Eqs. (\ref{motion2}) is considered by replacing $\frac{d\Omega_r}{dN}$, $ \frac{d\Omega_d}{dN}$ and $\frac{d\Omega_m}{dN}$ from Eqs. (\ref{system}). Now we can freely choose the two of them as independent variables. Here we take $\Omega _m$ and $\Omega _r$ as independent ones. So we will have $\Omega_d$ as a function of $\Omega _r$ and $\Omega_m$:
	
	\begin{equation}
	\label{od}
	\Omega_d=1-\Omega_m-\Omega_r.
	\end{equation}
	
	Thus replacing $\Omega _d$ by Eq. (\ref{od}),the equations for $\frac{d\Omega_m}{dN}$ and $\frac{d\Omega_r}{dN}$ would be:
	
	\begin{eqnarray}
	\label{dynamic1}
	&&\frac{d\Omega_m}{dN}=-3 \Omega_m -\Omega_m \gamma -\dfrac{A_2}{B_2},\\
	&&\frac{d\Omega_r}{dN}=-4 \Omega_r -\dfrac{A_3}{B_3},\label{dynamic2}
	\end{eqnarray}
	where:
	
	\begin{eqnarray}
	&&A_2=( 2 \Omega_m (9 \Omega_m^2 - 18 \Omega_m (1 - \Omega_m - \Omega_r) - 16 \Omega_r + \nonumber\\
	&&34 \Omega_m \Omega_r -  8 (1 - \Omega_m - \Omega_r) \Omega_r +24 \Omega_r^2 - \nonumber\\
	&& 3 \Omega_m (3 - 3 (1 - \Omega_m - \Omega_r) + \Omega_r) \gamma)),\nonumber\\
	\nonumber\\
	&&B_2=( 3 \Omega_m + 8 \Omega_r +   3 (-3 \Omega_m^2 +  \Omega_m (1 - \Omega_m - \Omega_r) +\nonumber\\
	&&4 (1 - \Omega_m - \Omega_r)^2 - 7 \Omega_m \Omega_r -4 \Omega_r^2 + 2 \Omega_m \gamma),\nonumber 
	\end{eqnarray}
	
	\begin{eqnarray}
	&&A_3=2 \Omega_r (9 \Omega_m^2 - 18 \Omega_m (1 - \Omega_m - \Omega_r) - 16 \Omega_r \nonumber\\
	&& + 34 \Omega_m \Omega_r -8 (1 - \Omega_m - \Omega_r) \Omega_r + 24 \Omega_r^2 - \nonumber\\
	&&3 \Omega_m (3 - 3 (1 - \Omega_m - \Omega_r) + \Omega_r) \gamma)),\nonumber\\
	\nonumber\\
	&&B_3=(3 \Omega_m + 8 \Omega_r +
	3 (-3 \Omega_m^2 + \Omega_m (1 - \Omega_m - \Omega_r) +\nonumber\\
	&&4 (1 - \Omega_m - \Omega_r)^2 - 7 \Omega_m \Omega_r -4 \Omega_r^2 + 2 \Omega_m \gamma).\nonumber
	\end{eqnarray}

	Using the linear dynamical systems theory, one can find the fixed points of the theory. However, before we proceed, it must be noted that the parameter that determine the stability of the fixed points is $\gamma$,  that is subject to a constraint that puts boundary on it. As $\Omega_r$, $\Omega_m$, and $\Omega_d$ are the energy portions of each energy content of the universe which should have positive values, we need to select values for $\gamma$ such that positive values are obtained for  $\Omega_r$, $\Omega_m$ and $\Omega_d$. This means that $\gamma$ can only be a positive constants. \\
	
	\subsection{Dynamical behavior of the universe by $f(R,T,R_{\mu \nu}T^{\mu \nu})$ theory }
	
\begin{figure} 
		\center
		\includegraphics[scale=0.6]{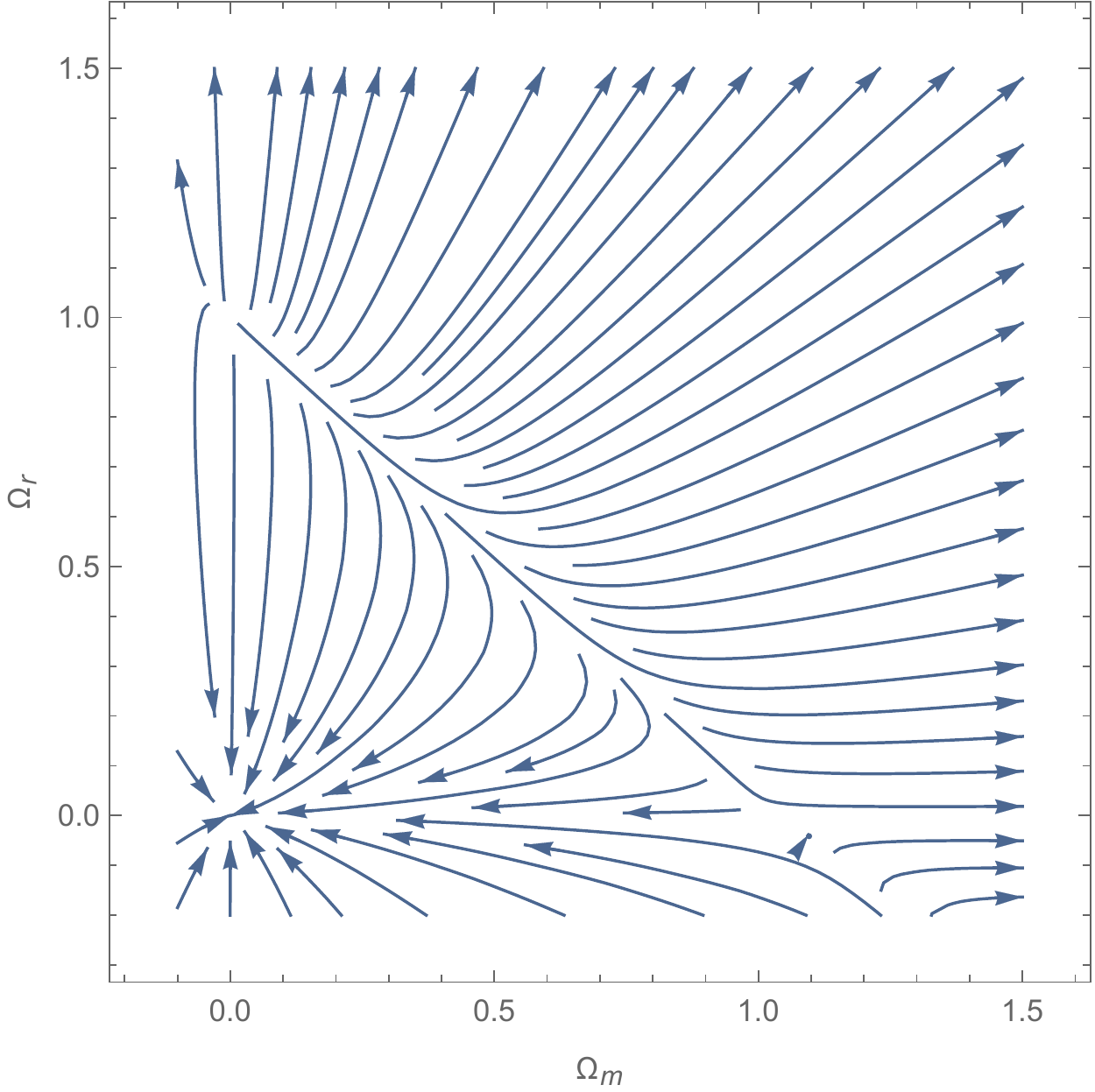}
		\caption{The behavior of phase trajectories of the system (Eqs.\eqref{dynamic1} and \eqref{dynamic2}) onto the phase plane ($\Omega_m,\Omega_r$) for $\gamma=0.1$, when we consider $\Omega_m$ and $\Omega_r$ as independent variables. The fixed points located on the positive region of the plot are considerable. The plot shows that $\Omega_m=0$ and $\Omega_r=1$ is an unstable fixed point as the flux is leaving this point toward the matter dominated fixed point.The point with $\Omega_m=1$ and $\Omega_r=0$ that shows the matter dominated universe is a saddle point. The flux leaves the radiation dominated universe to reach the matter dominate universe but it is a saddle point and it will continue to the stable fixed point ($\Omega_m=0$ and $\Omega_r=0$) in which the dark energy is dominated.}
		\label{fig:figure 8}
	\end{figure}

\begin{figure} 
		\center
		\includegraphics[scale=0.6]{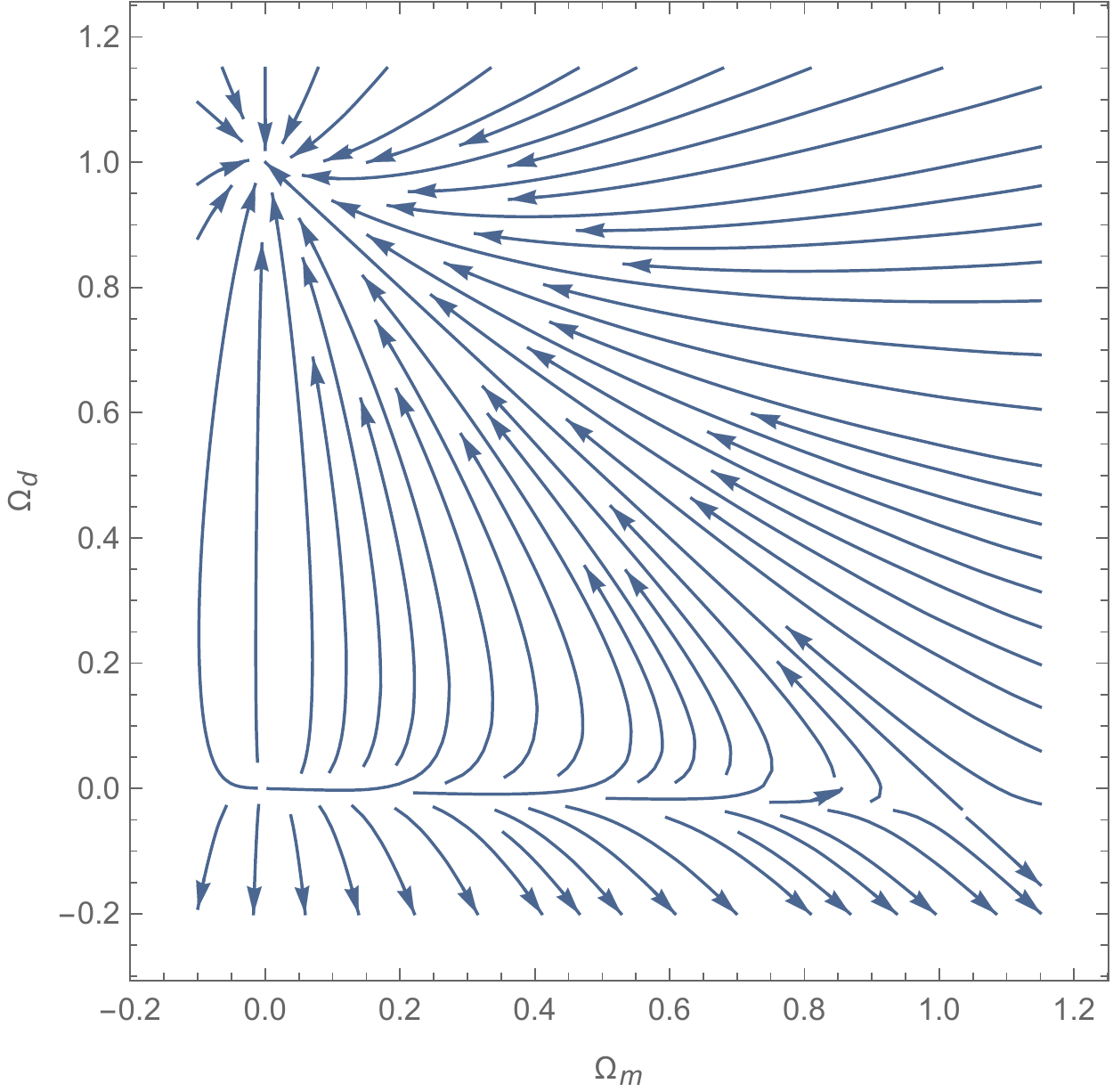}
		\caption{The behavior of phase trajectories of the system onto the phase plane ($\Omega_m,\Omega_d$) for $\gamma=0.1$, when we consider $\Omega_m$ and $\Omega_d$ as independent variables. The fixed points located on the positive region of the plot are considerable. The plot shows that $\Omega_m=0$ and $\Omega_d=0$ is an unstable fixed point that corresponds to the radiation dominated universe. The flux is leaving this point toward the matter dominated fixed point. The point with $\Omega_m=1$ and $\Omega_d=0$ that shows the matter dominated universe is a saddle point. The flux leaves the radiation dominated universe to reach the matter dominate universe but it is a saddle point and it will continue to the stable fixed point ($\Omega_m=0$ and $\Omega_d=1$) in which the dark energy is dominated. The results are consistent with that of the Fig.\ref{fig:figure 8}.}
		\label{fig:figure 9}
	\end{figure}

	Having selected the proper values, we are now in a position to check the fixed points of the theory. Eqs. \eqref{dynamic1} and \eqref{dynamic2} are the main equations for our present purpose. The result of using the linear dynamical system theory shows that there are three fixed points for our model: (${\Omega_m\rightarrow 0, \Omega_r\rightarrow1}$),  (${\Omega_m\rightarrow 1, \Omega_r\rightarrow0}$) and (${\Omega_m\rightarrow 0, \Omega_r\rightarrow0}$) which are representing a radiation dominated, matter dominated and dark energy dominated universe, respectively. The first one is where the eigenvalues of the Jacobian matrix are positive so it is an unstable fixed point and represents a universe that the radiation is dominant in it and will continue its evolution toward another epoch in which the matter is dominant. The second one (${\Omega_m\rightarrow 1, \Omega_r\rightarrow0}$) is a saddle point as expected. Because the universe flux is going toward this point from one side and leaves it from the other side to the next epoch. The third one (${\Omega_m\rightarrow 0, \Omega_r\rightarrow0}$) is where the dark energy is dominant. Considering the eigenvalues of the Jacobian matrix near this point, we find that it is a stable fixed point. This stability means that our universe tends to go toward a stable epoch in which the dark energy is dominant. The stability of these fixed points are shown in Fig.(\ref{fig:figure 8}). Also the dynamical behavior of the system when $\Omega _m$ and $\Omega_d$ are considered as independent variables instead $\Omega _m$ and $\Omega_r$, is depicted in Fig.(\ref{fig:figure 9}). The results are expected from the cosmological point of view. 
	The standard cosmological model can be summarized as follows \cite{fix}:\\
	inflation $\rightarrow$ radiation $\rightarrow$ matter $\rightarrow$ accelerating expansion\\
	A proposed \textcolor{black}{model should be consistent with at least a part} of the standard cosmological model discussed above. This means that there should exist a saddle matter fixed point or a saddle radiation fixed point, or an accelerated attractor (or all of them) in the theory.\\
	For the theory presented here, the stability of the dark energy era, and the instability of the radiation and the status of matter dominated universe era are shown in  Figs. (\ref{fig:figure 8}) and (\ref{fig:figure 9}) and are compatible with the above description.
	
	\section{Singularities of the $f=R+\alpha R_{\mu \nu}T^{\mu \nu}$ model}
	\label{IV}
	A future singularity for the universe is described as a divergence or vanishing of energy density, pressure, scale factor or Hubble parameter at a finite future time ($t_s$). Although there are some types of singularities which are defined in an infinite future time. The classification of the singularities is shown in Table \ref{class} \cite{4,6}. \\
	
	The occurrence of future singularities in the universe is highly sensitive to the gravitational models. Moreover the conditions in which the singularities occur may differ in various gravitational theories. Here to explore the future singularities, we need to solve  Eqs. (\ref{e}) and (\ref{f}) for $\rho$ and $p$.  Assuming the barotropic index ($w=\dfrac {p}{\rho}$) to be constant, one can solve these equations to find an expression for the energy density and pressure of the universe.\\

	\begin{table*}
		\caption{Classification of singularities in the FRW universe.}
		\label{class}
		\begin{tabular*}{1.05\textwidth}{@{\extracolsep{\fill}}lrrrrrrrrl@{}}
			\hline
			Type & Name & time & $a(t_s)$ & $\rho(t_s)$ & $p(t_s)$&$\dot{p}(t_s)$ etc. &$w(t_s)$& \\
			\hline
			$0$ & $Big-Bang(BB)$ & $0$ & 0 & $\infty $ & $\infty $ & $\infty $ & finite & \\
			$I$ & $Big-Rip(BR)$ & $t_s$ & $\infty $ & $\infty $& $\infty $& $\infty $ & finite & \\
			$I_{l}$ & $Little-Rip(LR)$ & $\infty$ & $\infty $ & $\infty $& $\infty $& $\infty $ & finite & \\
			$I_{p}$ & $Pseudo-Rip(PR)$ & $\infty$ & $\infty $ & finite & finite& finite& finite &\\
			$II$ & $Sudden\hspace{0.1cm} Future(SFS)$ & $t_s$ & $a_s$ & $\rho_{s}$& $\infty $& $\infty $& $\infty$ &\\
			$II_{g}$ & $Gen.\hspace{0.1cm} Sudden\hspace{0.1cm} Future(GSFS)$ & $t_s$ & $a_s$ & $\rho_{s}$& $p_s$& $\infty $& finite & \\
			$III$ & $Finite\hspace{0.1cm} Scale\hspace{0.1cm} Factor(FSF)$ & $t_s$ & $a_s$ & $\infty $& $\infty $& $\infty $& finite & \\
			$IV$ & $Big-Separation(BS)$ & $t_s$ & $a_s$ & 0&0& $\infty $& $\infty $ & \\
			$V$ & $w-Singularity(w)$ & $t_s$ & $a_s$ & 0&0&0& $\infty $ &\\
			\hline
		\end{tabular*}
	\end{table*}
	
	\textcolor{black}{We assume $H=H_0(t_s-t)^{-\beta}$ in which $H_0$ is the value of Hubble constant at present and is equal to $67.3 \pm 1.2 km\ s^{-1}MPc^{-1}$ \cite{1}, $t_s$ represents the time that future singularity occurs and the value of $\beta$ determines type of the future singularities, for Einstein-Hilbert action, as in Table \ref{beta} \cite{6,3b}.}
	
	\begin{table*}
		\caption{Types of future singularities for different values of $\beta$.}
		\label{beta}
		\begin{tabular*}{\textwidth}{@{\extracolsep{\fill}}lrrrrrrrrl@{}}
			\hline
			$\beta$ & Name & time & $a(t_s)$ & $\rho(t_s)$ & $p(t_s)$ \\
			\hline
			$\beta \geq 1$ & $ I $ & $t_s$ &$ \infty$ & $\infty $ & $\infty $   \\
			\hline
			$-1< \beta < 0$ & $II $ & $t_s$ & $a_s $ & $\rho _s $& $\infty $  \\
			\hline
			$0< \beta < 1$ & $ III$ & $t_s$ & $a_s $ & $\infty $& $\infty $ \\
			\hline
			$\beta \leq-1$ &$IV$   & $t_s$ & $a_s $ & $finite$ & $finite$ \\ 
			\hline
		\end{tabular*}
	\end{table*}
	It is obvious that by setting $p=w\rho$ with a constant value of $w$, we are allowed to check  just the occurrence of singularities of types I, III, and IV. For type II singularity, the energy density is finite but the pressure diverges; it will, therefor have no constant barotropic index. To simplify the calculations we assume that $\kappa=1$, $H_0=67 km\ s^{-1}MPc^{-1}$, $\alpha=\dfrac{1}{5}$ and $t_s=10$. \textcolor{black}{It should be noted that for smaller values of $\alpha$ the results are the same, however the time of singularity occurrence might be different.} \\
	\indent
	\noindent In the following subsections, the occurrence of singularities will be explored for, the four cases of $w=1$ (matter dominated universe), $w=-1$ (dark energy dominated universe), and $w=\dfrac{1}{3}$ (radiation dominated universe), and $w=0$ (dust matter dominated universe) in $f=R+\alpha R_{\mu \nu}T^{\mu \nu}$ model.

	\subsection{Radiation dominated universe}
	
	\label{sssec:3.1.1}
\begin{figure} 
		\center
		\includegraphics[scale=0.43]{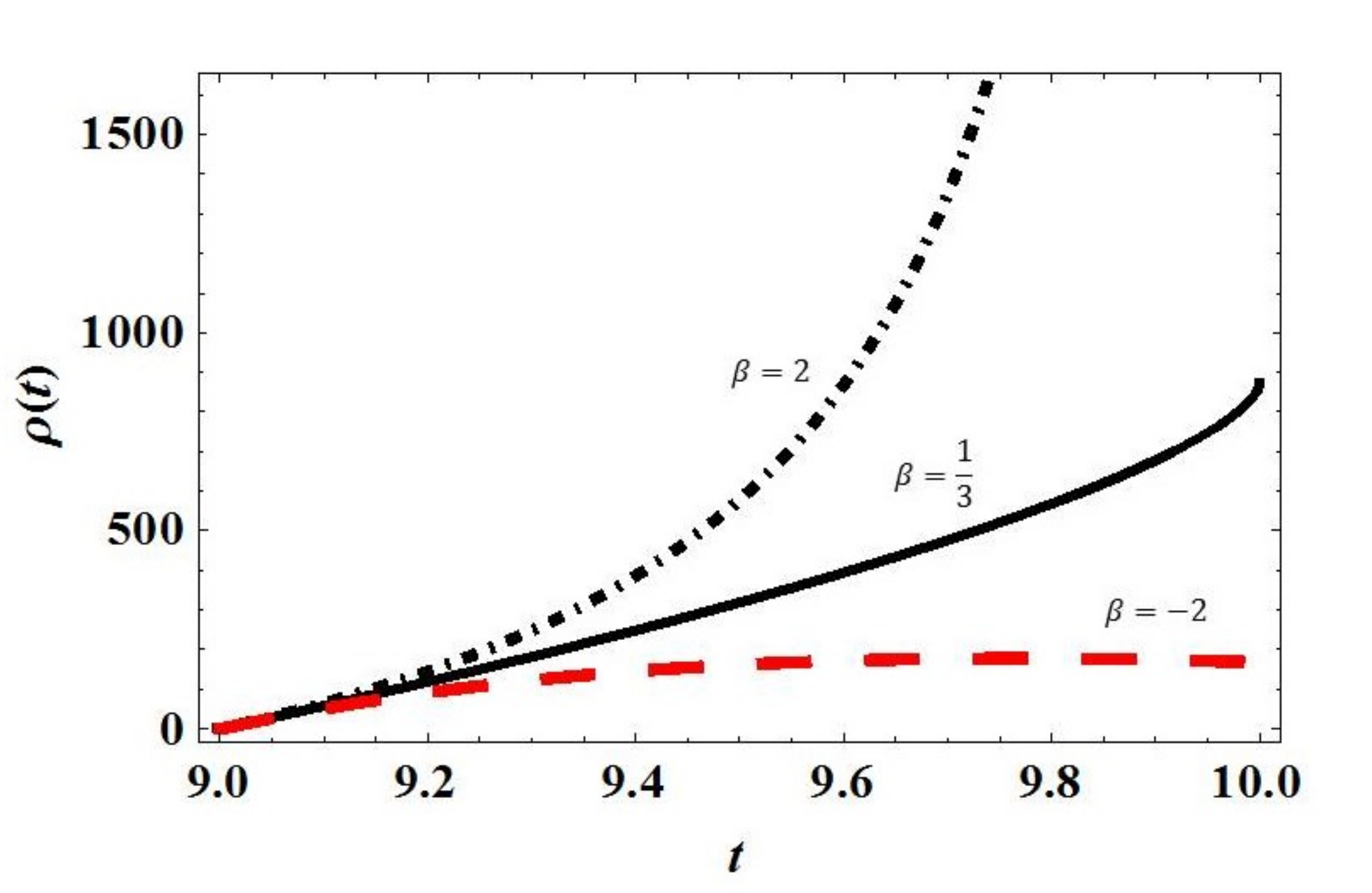}
		\caption{The energy density $\rho$, with respect to $t$ for $f=R+\alpha R_{\mu \nu}T^{\mu \nu}$ gravitational model with $w=\dfrac{1}{3}$, $\alpha=\dfrac{1}{5}$, $t_s=10$, and the three choices of $\beta=2$, $\beta=\dfrac{1}{3}$, and $\beta=-2$.}
		\label{fig:figure 1}
	\end{figure}

\begin{figure} 
		\center
		\includegraphics[scale=0.43]{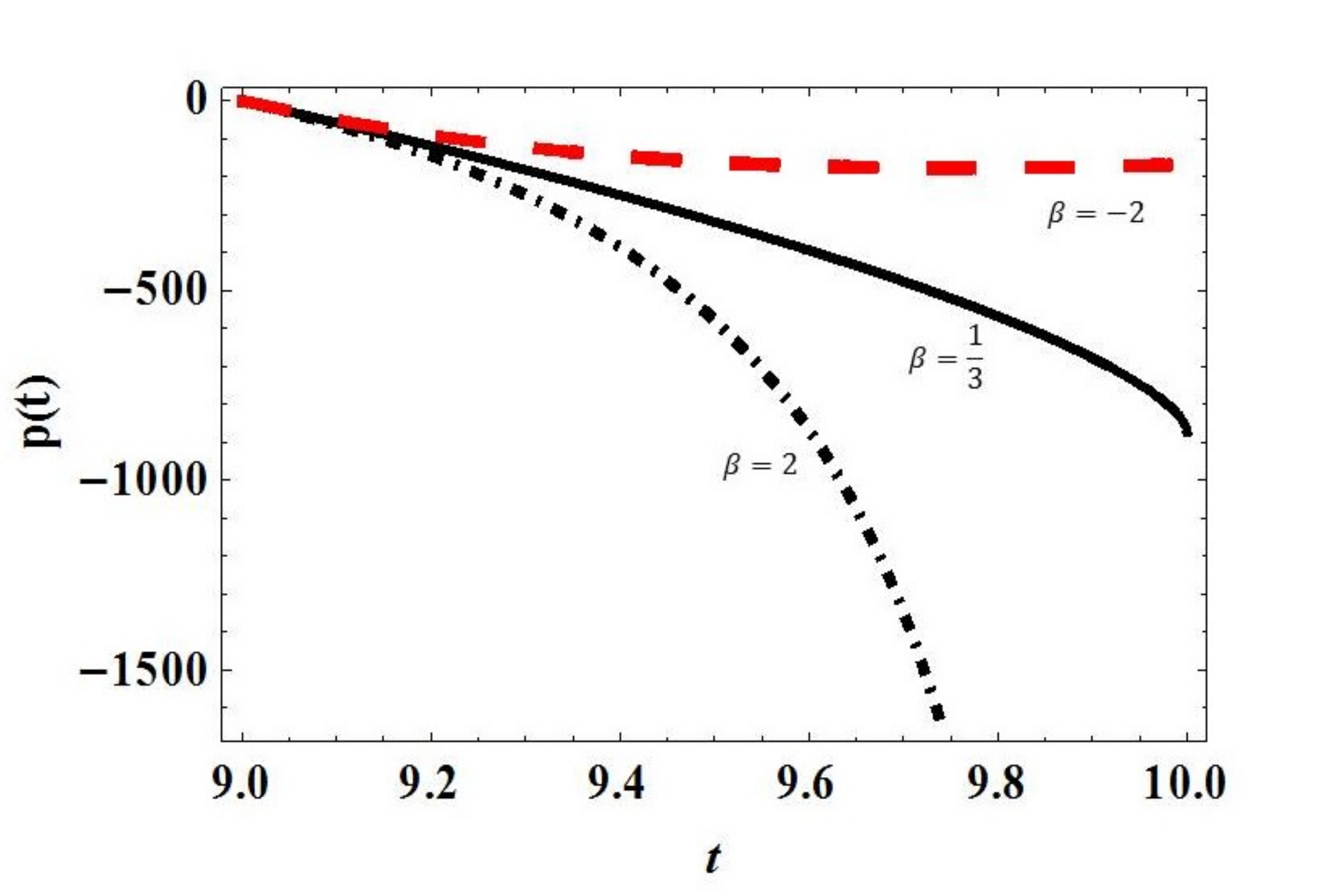}
		\caption{The pressure $p$, with respect to $t$ for $f=R+\alpha R_{\mu \nu}T^{\mu \nu}$ gravitational model with $w=\dfrac{1}{3}$, $\alpha=\dfrac{1}{5}$, $t_s=10$, and the three choices of $\beta=2$, $\beta=\dfrac{1}{3}$, and $\beta=-2$.}
		\label{fig:figure 2}
	\end{figure}

	\textcolor{black}{The barotropic index is $w=\dfrac{1}{3}$, in a radiation dominated universe}. Using $p=\dfrac{1}{3}\rho$ in Eqs. (\ref{e}) and (\ref{f}) for the three cases $\beta \geq 1$, $0< \beta < 1$ and $\beta \leq-1$, we will have two differential equations in terms of $\rho$ and it's derivatives. Figs. (\ref{fig:figure 1}) and (\ref{fig:figure 2}) present the numerical solutions of these equations representing  the behaviors of \textcolor{black}{the pressure and the energy density of the universe }for the three choices of $\beta$.

	For $\beta=2$, energy density and pressure tend to infinity at $t_s $  (i.e., the time we expect a singularity to occur). The scale factor and the Hubble parameter both diverge at $t_s$. Checking other values of  $\beta$ against $\beta \geq 1$ leads us to the conclusion that these values of $\beta$ give rise to a type I singularity. For $\beta=\dfrac{1}{3}$ ($0<\beta<1)$, the energy density, pressure, and the Hubble parameter tend to infinity at $t_s$; however, the scale factor is finite. Choosing $0< \beta < 1$ leads to a type III singularity. In the case of $\beta<-1$, the energy density and pressure have small  finite values around $t_s$ while the scale factor,  the Hubble parameter, and its first derivative are finite; so, $\beta \leq-1$ yields to a type IV singularity.
	

	\subsection{Dark Energy dominated universe with $w=-1$ }
	\label{sssec:3.1.2}
\begin{figure} 
		\center
		\includegraphics[scale=0.45]{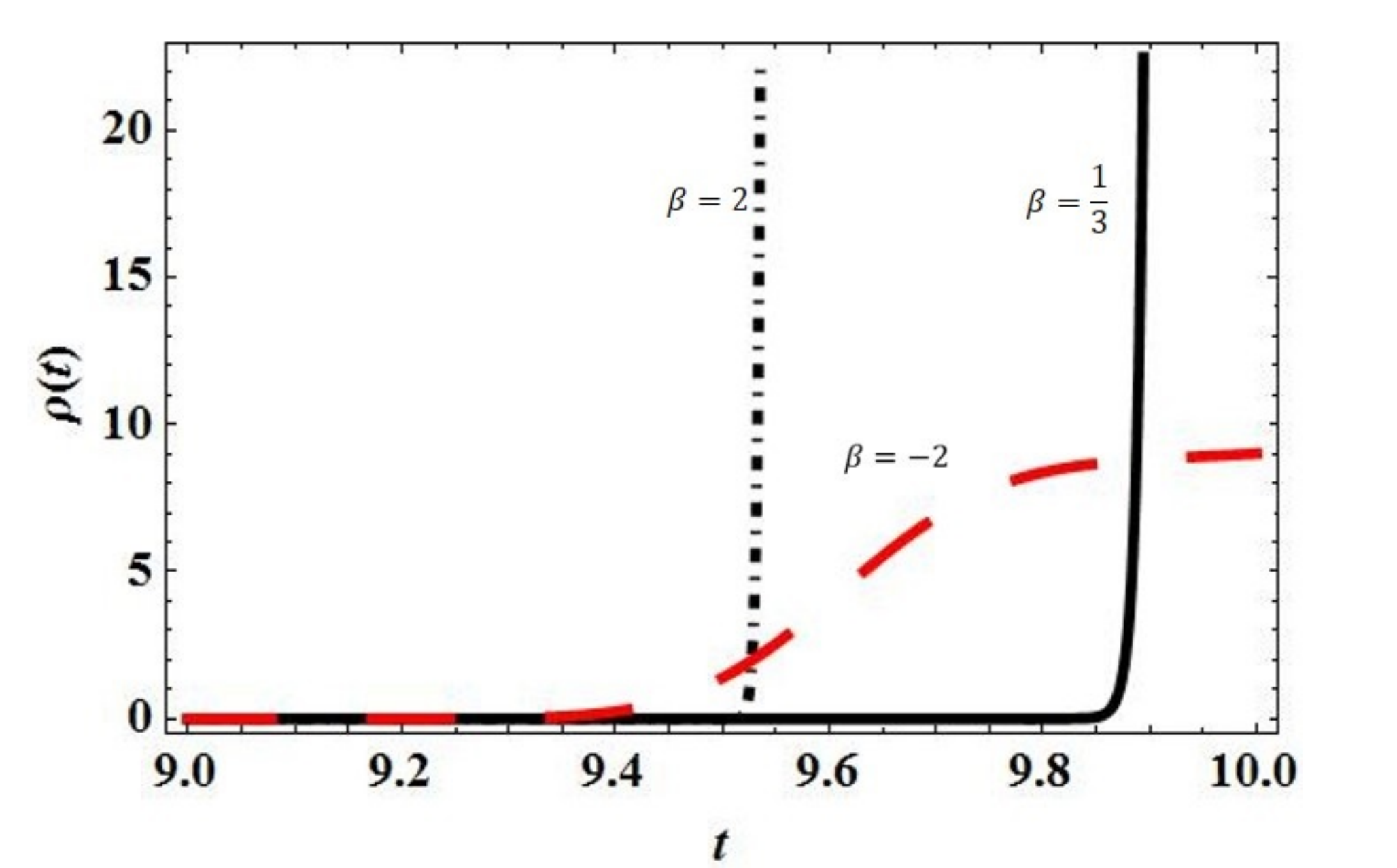}
		\caption{$\rho$ (energy density) with respect to $t$ (time) for $f=R+\alpha R_{\mu \nu}T^{\mu \nu}$ gravitational model with $w=-1$, $\alpha=\dfrac{1}{5}$, $t_s=10$, and the three choices of $\beta=2$, $\beta=\dfrac{1}{3}$, and $\beta=-2$.}
		\label{fig:figure 3}
	\end{figure}

\begin{figure} 
		\center
		\includegraphics[scale=0.45]{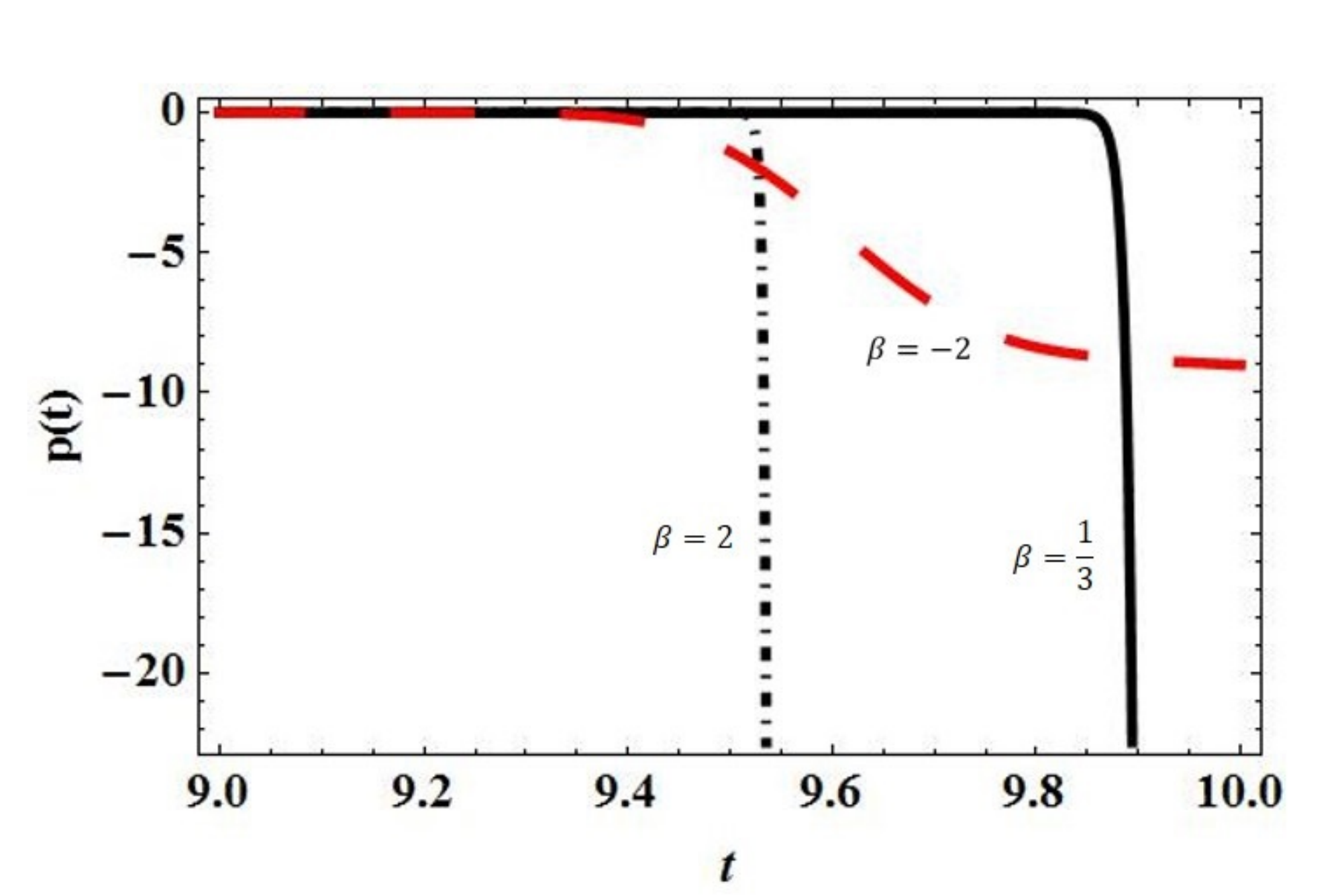}
		\caption{$p$ (pressure) with respect to $t$ (time) for $f=R+\alpha R_{\mu \nu}T^{\mu \nu}$ gravitational model with $w=-1$, $\alpha=\dfrac{1}{5}$, $t_s=10$, and the three choices of $\beta=2$, $\beta=\dfrac{1}{3}$, and $\beta=-2$.}
		\label{fig:figure 4}
	\end{figure}

	For a dark energy dominated universe with $p=-\rho$ $(w=-1)$ two differential equations are obtained in terms of p and its derivatives for the three cases of $\beta \geq 1$, $0< \beta < 1$, $\beta \leq-1$.  Numerical solutions of the energy density and pressure of the universe are depicted in Figs.(\ref{fig:figure 3}) and (\ref{fig:figure 4}) for the three choices $\beta=-2,\ \dfrac{1}{3}$ and, $2$.

	The results obtained for the dark energy case and the conditions for each singularity occurrence $(w=-1)$ are reported in Table \ref{DE}.

	\begin{table*}
		\caption{Singularities of the dark energy case with $w=-1$ and $\beta=-2,\dfrac{1}{2},\ 2$ for $f=R+\dfrac{1}{5} R_{\mu \nu}T^{\mu \nu}$ gravitational model.}
		\label{DE}
		\begin{tabular*}{\textwidth}{@{\extracolsep{\fill}}lrrrrrrrrl@{}}
			\hline
			& $\beta$ & $a(t_s)$ & $\rho(t_s)$ & $p(t_s)$ &$Singularity Type$& \\
			\hline
			& $-2\  (\beta <-1)$ & $a_s$ & $finite $ & $finite $ & $IV$ & \\
			& $\dfrac{1}{3} \ (0<\beta <1)$  & $a_s $ & $\infty $& $\infty $& $III$ & \\
			&$2 (\beta >1)\ $ & $\infty $ & $\infty $& $\infty $& $I $  & \\
			
			\hline
		\end{tabular*}
	\end{table*}


	\subsection{ High cosmological density limit of the field equation with $w=1$ }
	
	\label{sssec:3.1.3}
	For the high cosmological energy density limit in the field equations, the universe is viewed such that $p=\rho$. In addition , a small value is considered for the constant $\alpha$ so that $\alpha\rho<<1$ and $\alpha p<<1$. Effecting this approximation in Eqs. (\ref{e}) and (\ref{f}) will yield \eqref{g} and \eqref{h} below \cite{5a}:

	\begin{eqnarray}
	\label{g}
	3H^2&=&{\kappa}{\rho}, \\
	\label{h}
	2\dot{H}+3H^2&=&-{\kappa \rho}+2\alpha H\dot{\rho}.
	\end{eqnarray}
	
\begin{figure} 
		\center
		\includegraphics[scale=0.45]{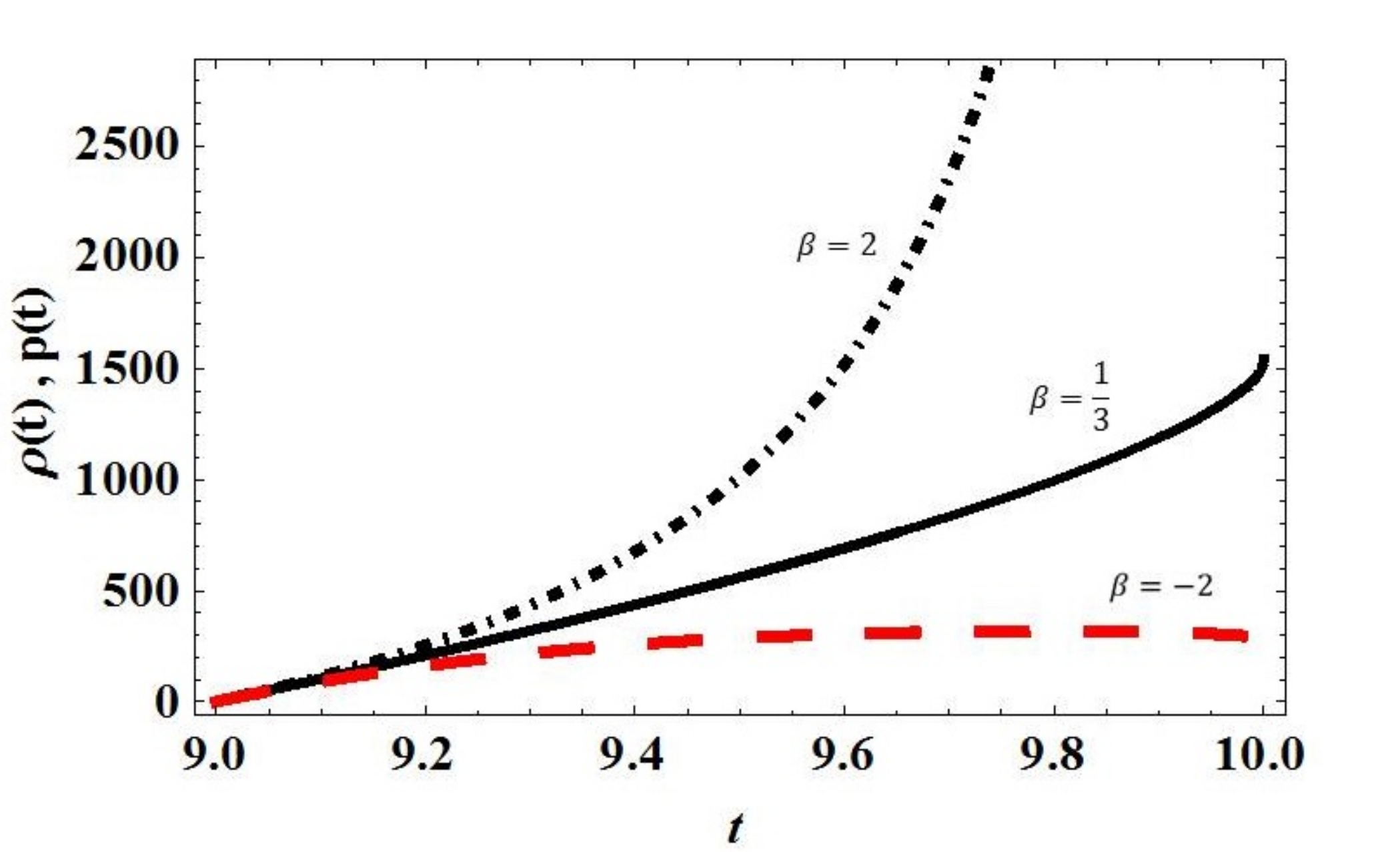}
		\caption{$p$ and $\rho$ with respect to $t$ (time) for $f=R+\alpha R_{\mu \nu}T^{\mu \nu}$ gravitational model with $w=1$, $\alpha=\dfrac{1}{5}$, $t_s=10$, and the three choices of $\beta=2$, $\beta=\dfrac{1}{3}$, and $\beta=-2$.}
		\label{fig:figure 5}
	\end{figure}

	Fig. (\ref{fig:figure 5}) illustrates the behaviors of energy dissipation and pressure for $\beta \geq 1$, $-1< \beta < 0$ and $\beta \leq-1$. Table \ref{HC} provides the information about the  singularity types pertinent to the  high cosmological energy density limit.

	\begin{table*}
		\caption{Singularities of the high cosmological limit for the field equations with $w=1$ and $\beta=-2,\ \dfrac{1}{2},$ and $\ 2$ for $f=R+\dfrac{1}{5} R_{\mu \nu}T^{\mu \nu}$ gravitational model.}
		\label{HC}
		\begin{tabular*}{\textwidth}{@{\extracolsep{\fill}}lrrrrrrrrl@{}}
			\hline
			& $\beta$ & $a(t_s)$ & $\rho(t_s)$ & $p(t_s)$ &$Singularity Type$& \\
			\hline
			& $-2 (\beta <-1)$ & $a_s$ & $finite $ & $finite $ & $IV$ & \\
			& $\dfrac{1}{3} (0<\beta <1)$  & $a_s $ & $\infty $& $\infty $& $III$ & \\
			&$2 (\beta >1)$ & $\infty $ & $\infty $& $\infty $& $I $  & \\
			
			\hline
		\end{tabular*}
	\end{table*}
	
	\subsection{The case of dust matter with $w=0$}
	\label{3-1}
	
	
	We now turn to the case in which the density of cosmological matter is very low with $p=0$ ($w=0$). We assume again that $\alpha\rho\ll1$. The approximate forms of Eqs. (\ref{e}) and (\ref{f}) will then take the following forms \cite{5a}:

	\begin{eqnarray}
	&&3H^2={\kappa}{\rho}+\frac{3}{2}{\alpha}H\dot {\rho},\label{i}\\
	&&2\dot{H}+3H^2=2\alpha H\dot{\rho}+\frac{1}{2}{\alpha}\ddot{\rho}.\label{j}
	\end{eqnarray}

\begin{figure} 
		\center
		\includegraphics[scale=0.45]{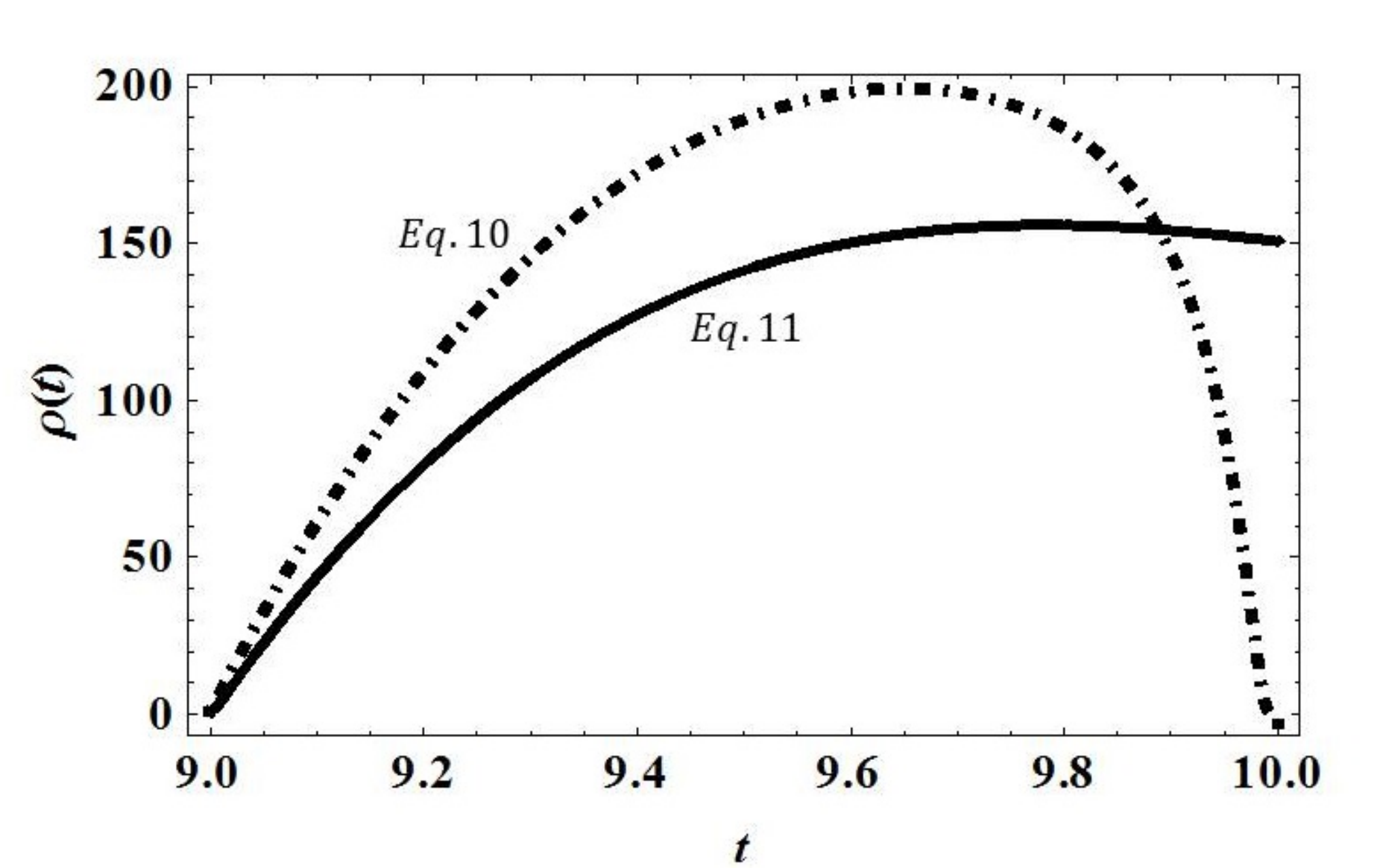}
		\caption{$\rho$ with respect to $t$ (time) for $f=R+\alpha R_{\mu \nu}T^{\mu \nu}$ gravitational model with $w=0$, $\alpha=\dfrac{1}{5}$, $t_s=10$, and the three choices of $\beta=2$, $\beta=\dfrac{1}{3}$, and $\beta=-2$.}
		\label{fig:figure 6}
	\end{figure}

	\noindent Pressure is equal to zero for a cosmological matter, and only a type IV singularity is expected. It should be noted that $p$ has an infinite value in the other three cases, as  shown in Part II, but $p\rightarrow0$ in type IV.
	In this case, the values of $\kappa$, $\alpha$ in Eqs. (\ref{i}) and (\ref{j}) may be employed for the case of ($\beta \leq-1$), to obtain the values for $\rho$ from Eq.(\ref{i}) and Eq.(\ref{j}) separately. Fig. (\ref{fig:figure 6}) illustrates the behavior of energy density for Eqs. (\ref{i}) and (\ref{j}).

	Clearly, $\rho$ has a finite value around $t_s$ so that choosing $\beta \leq-1$ yields a type IV singularity.

	It is evident from the diagrams in Figs. (\ref{fig:figure 1})-(\ref{fig:figure 6}), that finite future singularities in the $f=R+\alpha R_{\mu \nu}T^{\mu \nu}$ gravitational model appear to be the same as the Einstein-Hilbert gravity.

	\section{Field Equations and singularities of the $f=R + \alpha R R_{\mu \nu}T^{\mu \nu}$ model}
	\label{V}
	
	\indent
	In this Section, we study the cosmological model captured by  $f(R, T, R_{\mu\nu}T^{\mu\nu})=R + \alpha R R_{\mu \nu}T^{\mu \nu}$. Replacing this function in the action Eq. (\ref{a}) and  assuming pressure to be equal to zero (dust matter) for the matter filled universe, the following field equation obtained:
	
	\begin{eqnarray}
	\label{second}
	&&\bigg[1+\alpha(R_{\alpha\beta}T^{\alpha\beta}-RL_m)\bigg]G_{\mu\nu}+\alpha\bigg[\Box(R_{\alpha\beta}T^{\alpha\beta})
	+\nabla_\alpha\nabla_\beta(RT^{\alpha\beta})\bigg]g_{\mu\nu}-\nonumber\\
	&&\alpha\nabla_\mu\nabla_\nu(R_{\alpha\beta}T^{\alpha\beta})
	+\dfrac{1}{2}\alpha\Box(RT_{\mu\nu})+2\alpha RR_{\alpha(\mu}T^\alpha_{~\nu)}-\alpha\nabla_\alpha\nabla_{(\mu}\big[RT^\alpha_{~\nu)}\big]\nonumber\\
	&&-\bigg(\dfrac{1}{2}\alpha R^2+8\pi G\bigg)T_{\mu\nu}-2\alpha RR^{\alpha\beta}\dfrac{\partial^2 L_m}{\partial g^{\mu\nu}\partial g^{\alpha\beta}}=0.
	\end{eqnarray}
	
	Assuming the FRW metric the cosmological field equations will take the form as bellow \cite{5a}:
	
	\begin{align}
	\label{energy}
	&-3H^2 + \rho+\alpha(18H\ddot{H}\rho + 18H
	\dot{H}\dot{\rho} + 54H^2\dot{H}\rho - \nonumber\\
	&9\dot{H}^2\rho + 27H^3\dot{\rho} +
	27H^4\rho) = 0,
	\end{align}
	
	\begin{align}
	\label{pressure}
	&-2\dot{H} - 3H^2 + \alpha(6\dddot{H}\rho + 12\ddot{H}\dot{\rho} +
	36H\ddot{H}\rho + \nonumber\\
	& 6\dot{H}\ddot{\rho} +54H\dot{H}\dot{\rho} + 48H^2\dot{H}\rho +
	15\dot{H}^2\rho + \nonumber\\
	&9H^2\ddot{\rho} +30H^3\dot{\rho} - 9 H^4\rho) = 0.
	\end{align}
	
\begin{figure} 
		\center
		\includegraphics[scale=0.44]{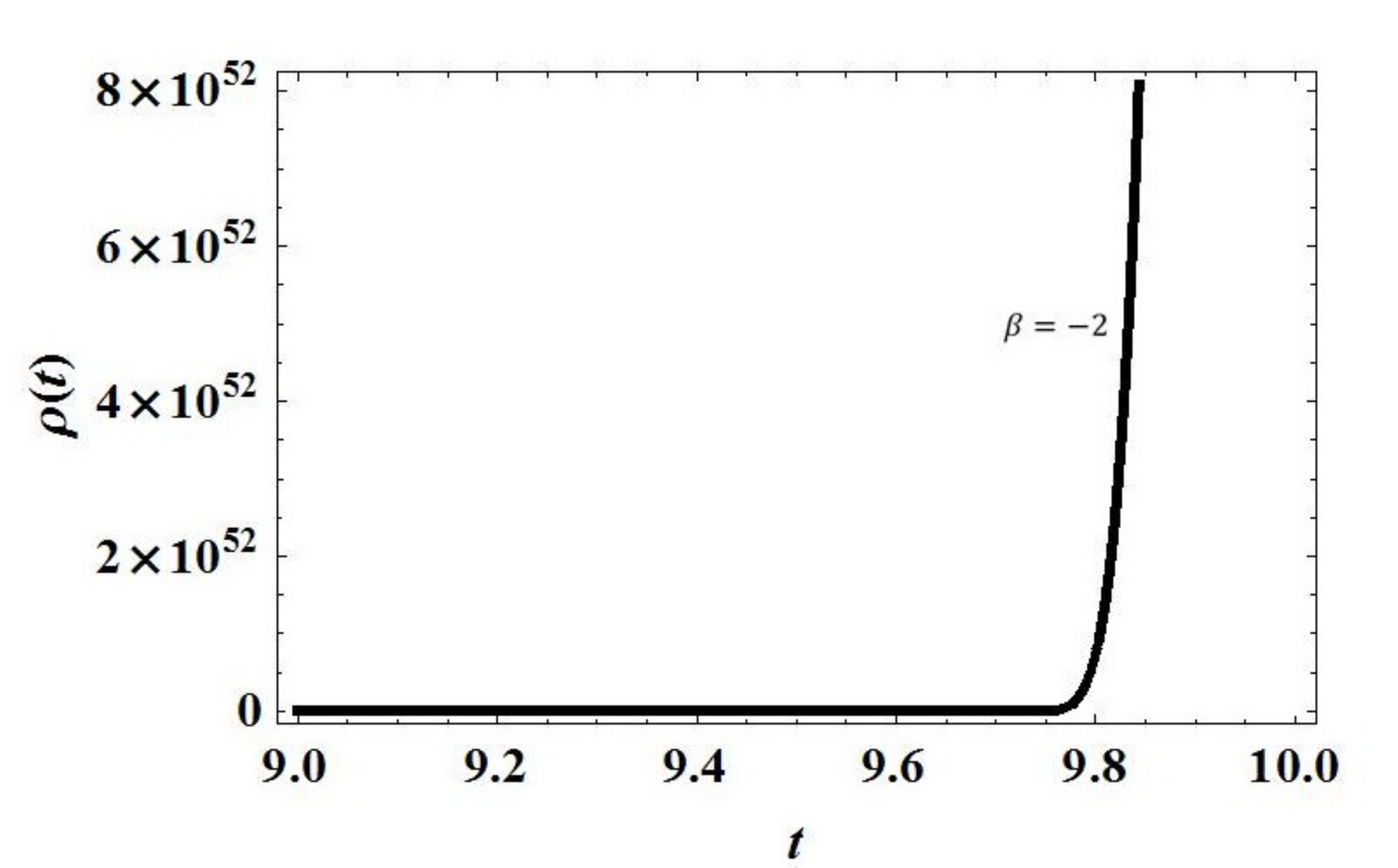}
		\caption{$\rho$ with respect to $t$ (time) for $f=R+\alpha RR_{\mu \nu}T^{\mu \nu}$ gravitational model with $w=0$, $\alpha=\dfrac{1}{5}$, $t_s=10$, and $\beta=2$.}
		\label{fig:figure 7}
	\end{figure}

	As for the dust matter, the barotropic index is equal to zero. Solving Eqs. (\ref{energy}) and (\ref{pressure}) for the energy density will yield two differential equations for $\rho$ that must be equal to each other. This equality gives us a unique equation which can be solved numerically and the behavior of the energy density of the universe in this model will be the one shown in the Fig.(\ref{fig:figure 7}). The parameter $\beta=-2$ corresponds to the case that predict a type IV singularity in the FRW universe with an Eintein-Hilbert action for the future of the universe (in which, the scale factor, the Hubble parameter, \textcolor{black}{the pressure, and the energy density are nearly zero while also the Hubble parameter's higher derivatives diverge at the time of singularity, $t_s$)}. This is while in the cosmological model studied above revealed that in here, for $\beta =-2$, the energy density of the universe is going to diverge at the time of singularity while the scale factor, the Hubble parameter, and the pressure still remain to be nearly zero. To resolve this, a new type of singularity with the following properties may be introduced:
	
	\begin{equation}
	\label{new}
	t\rightarrow t_s,\  a\rightarrow a_s,\  \rho \rightarrow \infty,\ and  \  |p| \rightarrow 0 \  (\beta \leq-1).
	\end{equation}
	
	\indent
	For the other values of $\beta$ and barotropic index, the behaviors of the energy density and the pressure of the universe are the same as those described at the beginning of Sec. \ref{IV}.

\section{Cosmographic Study}
\label{V1}
In this section we are going to do some numerical studies on $f=R+\alpha R_{\mu \nu}T^{\mu \nu}$ model using cosmographic parameters. In this method the main idea is to determine the value of the free parameters of a certain gravitational theory using the derivatives of scale factor\cite{cg1}. The cosmographic parameters are defined by derivatives of the scale factor as follows:

\begin{eqnarray}
&&H(t)=\dfrac{\dot{a}(t)}{a(t)}\\
&&q(t)=\dfrac{-1}{a(t)}\ddot{a}(t)H(t)^{-2}\\
&&j(t)=\dfrac{-1}{a(t)}\dddot{a}(t)H(t)^{-3}\\
&&s(t)=\dfrac{-1}{a(t)}\ddddot{a}(t)H(t)^{-4}\\
&&l(t)=\dfrac{-1}{a(t)}a(t)^{(5)}H(t)^{-5}\\
&&m(t)=\dfrac{-1}{a(t)}a(t)^{(6)}H(t)^{-6}
\end{eqnarray}

\noindent which are Hubble, deceleration, jerk, snap, lerk and m parameters, where their present values are known observationally as $H_0$,$q_0$,$j_0$,$s_0$. To determine the free parameters ($\rho_r,\ \rho_m,\ \rho_d,\ \alpha$ and $\gamma$), in a way that they are compatible with observation, we must take the following path \cite{cg1}:
 \begin{enumerate}
 	\item The first Friedmann equation should be written in terms of scale factor and its derivatives (Eq. \eqref{friedmann}).
 	\item Using the conservation equations in Eqs. \eqref{conserv} to replace the time derivatives of energy densities as a function of the scale factor.
 	\item Making derivatives of the first Friedmann equation as many times as the number of free parameters.
 	\item Replacing the derivatives of scale factor with cosmographic parameters  ($H_0,\ q_0, j_0,\ s_0$).
 	\item Solving the obtained equations for the free parameters.
 	\item Find the values of the free parameters.
 \end{enumerate}
Here, we have five free parameters to be determined: $\rho_r,\ \rho_m,\ \rho_d,\ \alpha$ and $\gamma$, which for the first three we will find their present values. Note that the current value of radiation energy density ($\rho_r$) is much lower than the other components of energy densities, so we neglect its present value in our calculation. Therefore, the first three derivatives of Friedmann equation is required for the procedure. \\
We use the sets of data for cosmographic parameters \cite{cg2} and solve the algebraic equations for $ \rho_m,\ \rho_d,\ \alpha$ and $\gamma$.\\

At first we use the set of data from SNIa  Hubble diagram ,BAO and $H(z)$ in Table \ref{tab1}\cite{cg2} and our results for free parameters are presented in Table \ref{tab11}. Instead of $\rho_m$ and $\rho _d$, we have used $\Omega_m=\frac{\rho_m}{3H^2}$ and $\Omega_d=\frac{\rho_d}{3H^2}$ in Tables \ref{tab11} and \ref{tab22}.

\begin{table}
	\begin{center}
		\centering
		\setlength{\tabcolsep}{0.1 em}
		\begin{tabular}{|c|c|c|c|c|c|}
			
			\hline
			Parameter&$H_0$&$q_0$&$j_0$&$s_0$\\
			\hline
			\hline
			Best fit & $74$&$-0.48$ & $0.68$ & $-0.51$ \\
			Mean & $74$&$ -0.48$&$0.65$&$-6.8$ \\
			2 $\sigma$ & $ (68, 72)$&$ (-0.5, -0.38)$&$(0.29, 0.98)$&$(-1.33, -0.53)$\\
			\hline
		
		\end{tabular}
	\end{center}
	\caption{cosmographic parameters using the data from SNIa  Hubble diagram ,BAO and $H(z)$ \cite{cg2}.}
	\label{tab1}
\end{table}

\begin{table}
	\begin{center}
		\centering
		\setlength{\tabcolsep}{0.1 em}
		\begin{tabular}{|c|c|c|c|c|c|}
			
			\hline
			Parameter&$\Omega_m$&$\Omega_d$&$\alpha$&$\gamma$\\
			\hline
			\hline
			value & $0.11$&$0.6$ & $8\times 10^{-5}$ & $(-5,0.4)$ \\
			\hline
			
		\end{tabular}
	\end{center}
	\caption{Free parameters of the theory using cosmographic parameters obtained from the SNIa  Hubble diagram ,BAO and $H(z)$ data sets.}
	\label{tab11}
\end{table}

The cosmographic parameters obtaind using the set of data from GRBs Hubble diagram, BAO and $H(z)$ are presented in Table \ref{tab2}\cite{cg2}, and our results for free parameters are presented in Table \ref{tab22}. 

\begin{table}
	\begin{center}
		\centering
		\setlength{\tabcolsep}{0.1 em}
		\begin{tabular}{|c|c|c|c|c|c|}
		
			\hline
			Parameter&$H_0$&$q_0$&$j_0$&$s_0$\\
			\hline
			\hline
			Best fit & $67$&$-0.14$ & $0.6$ & $-5.55$ \\
			Mean & $67$&$ -0.14$&$0.6$&$-5.55$ \\
			2 $\sigma$ & $ (66, 73)$&$ (-0.15, -0.14)$&$(0.58,0.62)$&$(-5.7,6.1)$\\
			\hline
			
		\end{tabular}
	\end{center}
	\caption{cosmographic parameters  using the GRBs Hubble diagram , BAO and $H(z)$ data sets \cite{cg2}.}
	\label{tab2}
\end{table}

\begin{table}
	\begin{center}
		\centering
		\setlength{\tabcolsep}{0.1 em}
		\begin{tabular}{|c|c|c|c|c|c|}
			
			\hline
			Parameter&$\Omega_m$&$\Omega_d$&$\alpha$&$\gamma$\\
			\hline
			\hline
			Value & $0.266$&$0.65$ & $8\times 10^{-5}$ & $(-7,-2.9)$ \\
			\hline
			
		\end{tabular}
	\end{center}
	\caption{Free parameters of the theory using cosmographic parameters obtained from the GRBs Hubble diagram , BAO and $H(z)$  data sets.}
	\label{tab22}
\end{table}

The results in Tables \ref{tab11} and \ref{tab22} shows that the $f=R+\alpha R_{\mu \nu}T^{\mu \nu}$ model is consistent with the observational results for matter and dark energy. In addition as expected, cosmographic analysis of this model shows that the parameters $\alpha$ has a very small value and $\gamma$ is negative. 
	
\section{Conclusion}
\label{VI}
	
Different aspects of  $f(R, T, R_{\mu\nu}T^{\mu\nu})$ gravitational theory has already been studied in the literature \cite{5a,S1,S11,S12,S13}. The results indicate that $f(R, T, R_{\mu\nu}T^{\mu\nu})$ gravity is a consistent reliable cosmological model. In this theory, the late time acceleration of the universe can be considered as a result of matter-geometry non-minimal coupling. \\
In this work the method for deriving an autonomous dynamical system for the special case of $f=R+\alpha R_{\mu \nu}T^{\mu \nu}$ was explored to find radiation, matter and dark energy dominated universe in the theory. For this purpose we have construct an appropriate dynamical system equations which are made dimensionless using $H^2$ as a normalization parameter. We have assumed an interaction between matter and dark energy by the factor of $\gamma$.  Autonomous dynamical system method shows that the system has three fixed points which represent radiation, matter and dark energy dominated respectively. It was also found that while the fixed point for radiation dominated was unstable for $\gamma=0.1$, $\Omega_r=1,\Omega_m=0$, matter dominated one was of the saddle and that transition from the radiation era to the matter one was possible. Finally, it was observed that the dark energy dominated era was a stable fixed point for $\Omega_r=0$, and $\Omega_m=0$ and positive $\gamma$. These results are depicted in Figs. (\ref{fig:figure 8}-\ref{fig:figure 9}).\\
	
We also studied the occurrence of future singularities of the universe in the $f(R, T, R_{\mu\nu}T^{\mu\nu})$ gravity. In particular, two forms of this theory (i.e., $f=R+\alpha R_{\mu \nu}T^{\mu \nu}$ and $f=R+\alpha RR_{\mu \nu}T^{\mu \nu}$) were studied.\\
\indent The occurrence of the future singularities of the universe are predictable in many gravitational and cosmological theories such as the standard FRW model. These singularities are described by divergence or vanishing of one or some of the parameters $\rho, p, a, H$ and higher derivatives of the Hubble parameter. \textcolor{black}{ We assume the hubble parameter as $h=H_0(t-t_s)^{-\beta}$, where $\beta$ is a constant that can determine the type of singularity which may occur.} Efforts were made in this study  to check if the future singularities would occur in the $f(R, T, R_{\mu\nu}T^{\mu\nu})$ theory and if so under what conditions. Applying the field equations of the first case for the three different values of $\beta$ and for $w=1, -1,\dfrac{1}{3}$ and $w=0$, revealed that the relationships between the values of $\beta$ and the types of future singularities were the same as those under the FRW gravity (Figs. \ref{fig:figure 1}, \ref{fig:figure 2}, \ref{fig:figure 3}, \ref{fig:figure 4}, \ref{fig:figure 5}, \ref{fig:figure 6}). In the case $f=R+\alpha RR_{\mu \nu}T^{\mu \nu}$, however, the study of field equations and the singularities which might happen in the future of the universe unfolded interesting result. More specifically, it was found that, for $\beta \leq - 1$ and $w=0$, which belongs to type IV singularity in the FRW model, new type of singularity might be assumed to exist in the case of $F=R+\alpha RR_{\mu \nu}T^{\mu \nu}$ that was described by $t\rightarrow t_s,\  a\rightarrow a_s,\ \rho \rightarrow \infty,\ $  and $\  |p| \rightarrow 0\  (\beta \leq-1)$. The reason is the presence of an explicit coupling between matter and geometry in tensorial part, by the factor of $\alpha$. The inclusion of the time derivatives of pressure and energy density due to the coupling between $R_{\mu \nu}$ and $T^{\mu\nu}$ makes the singularity type to be different from those in GR. This is interesting to find universality classes of the future singularities in modified gravitational theories.\\
Finally a cosmographic analysis on this theory shows that for the theory to be consistent with observational data $\alpha$ should be of the order of $10^{-5}$ and $\gamma$ needs to be negative (and about $-2.9$).

\section*{Acknowledgment}
We thank Tiberiu Harko and Gonzalo Olmo for reading this paper, useful comments and suggestions. This work has been supported financially by Research Institute for Astronomy and Astrophysics of Maragha (RIAAM) under research project No. 1/4717-73.

\bibliographystyle{abbrv}
\bibliography{arxiv_version}

\end{document}